\documentclass[useAMS,usenatbib]{mn2e}
\usepackage{epsfig,graphicx,lscape}
\title[Fe XIII lines in SERTS spectra]{Fe XIII emission lines in active region spectra obtained with the 
Solar Extreme-Ultraviolet Research Telescope and Spectrograph}

\author[F. P. Keenan et al.]{F. P. Keenan\thanks{E-mail:
F.Keenan@qub.ac.uk},$^{1}$ D. B. Jess,$^{1}$ K. M. Aggarwal,$^{1}$
R. J. Thomas,$^{2}$ J. W. Brosius$^{2,3}$ \newauthor
and J. M. Davila$^{2}$
\\
$^{1}$Astrophysics Research Centre, School of Mathematics and Physics, Queen's University, Belfast BT7 1NN
\\
$^{2}$Laboratory for Solar Physics, Code 671, Heliophysics Science Division,
NASA's Goddard Space Flight Center, Greenbelt, 
\\
MD 20771, USA
\\
$^{3}$Department of Physics, The Catholic University of
America, Washington, DC 20064, USA
}

\begin{document}

\date{Accepted 2006 XXXX. Received 2006 XXXX; in original form 2006 XXXX}

\pagerange{\pageref{firstpage}--\pageref{lastpage}} \pubyear{}

\maketitle

\label{firstpage}

\begin{abstract}
Recent fully relativistic calculations of radiative rates and electron impact
excitation cross sections for Fe\,{\sc xiii} are used to generate
emission-line ratios involving 3s$^{2}$3p$^{2}$--3s3p$^{3}$ and 
3s$^{2}$3p$^{2}$--3s$^{2}$3p3d transitions in the 170--225\,\AA\ and
235--450\,\AA\ wavelength ranges covered by the 
Solar Extreme-Ultraviolet Research Telescope
and Spectrograph (SERTS). A comparison of these line ratios
with SERTS active region observations from rocket flights in 1989 and 1995
reveals generally very good agreement between theory and
experiment. Several new Fe\,{\sc xiii}
emission features are identified, at wavelengths of 203.79,
259.94, 288.56 and 290.81\,\AA.
However, major discrepancies between theory and observation 
remain for several Fe\,{\sc xiii}
transitions, as previously found by Landi (2002) and others, which cannot be explained
by blending. Errors in the adopted atomic data appear to be the most likely
explanation, in particular for transitions which have 3s$^{2}$3p3d $^{1}$D$_{2}$ 
as their upper level. 
The most useful Fe\,{\sc xiii}
electron density diagnostics
in the SERTS spectral regions are 
assessed, in terms of the line pairs involved being
(i) apparently free of atomic physics problems and blends, (ii)
close in wavelength to
reduce the effects of possible errors in the instrumental 
intensity calibration, and (iii)
very sensitive to changes in N$_{e}$ over the range 10$^{8}$--10$^{11}$\,cm$^{-3}$.
It is concluded that the ratios which best satisfy 
these conditions are 200.03/202.04 and 203.17/202.04 for the 170--225\,\AA\ wavelength region, 
and 348.18/320.80, 348.18/368.16, 359.64/348.18 and 359.83/368.16
for 235--450\,\AA.

\end{abstract}


\section{Introduction}

This is the latest in a series of papers in which we compare
theoretical emission-line intensity ratios, generated using accurate 
atomic physics data for energy levels, radiative rates, and electron impact 
excitation cross sections, with high resolution spectra from the Solar 
Extreme-Ultraviolet Research Telescope and Spectrograph (SERTS).
The major aim of our work is to investigate 
the importance of blending in the solar extreme-ultraviolet
spectral regions covered by SERTS (170--225\,\AA; 235--450\,\AA),
and hence determine which emission lines
provide the most reliable
diagnostics, and also where possible to identify new 
emission features.
To date, we have 
studied the SERTS
spectra of C\,{\sc iv} (Keenan et al. 1993), Mg\,{\sc ix} (Keenan et al. 1994),
Si\,{\sc x} (Keenan et al. 2000a), S\,{\sc xi} (Keenan et al. 2000b), 
Mg\,{\sc vi} (Keenan et al. 2002),
Si\,{\sc ix} (Keenan et al. 2003a),
Si\,{\sc viii} (Keenan et al. 2004), Na-like ions between 
Ar\,{\sc viii} and Zn\,{\sc xx} (Keenan et al. 2003b), and the iron ions
Fe\,{\sc xi} (Pinfield et al. 2001; Keenan et al. 2005a), Fe\,{\sc xii} 
(Keenan et al. 1996) and Fe\,{\sc xv} (Keenan et al. 2005b).

Our work is currently of particular relevance due to the recent launch
of the {\em Hinode} mission. This has on board the EUV Imaging
Spectrometer (EIS), which obtains high spectral resolution
observations over wavelength ranges similar to those covered by SERTS,
specifically 170--210\,\AA\ and 250--290\,\AA\ (Culhane et al. 2005). 
Clearly, it is important that the emission lines observed by
the EIS are investigated in detail and the best diagnostics identified,
with SERTS providing the ideal testbed for this. In the present paper we undertake a
study of Fe\,{\sc xiii}, which as noted by Young, Landi \& Thomas (1998)
is the species with the most emission lines detected in the SERTS spectral
range. It has also long been known that Fe\,{\sc xiii}
emission lines provide very useful electron density diagnostics for solar
plasmas (Flower \& Nussbaumer 1974).

The most detailed investigation to date 
of the Fe\,{\sc xiii}
solar extreme-ultraviolet spectrum, which also employs SERTS observations, is probably that 
by Landi (2002). In addition, his paper contains an excellent review of
previous work on Fe\,{\sc xiii}. Landi undertook two separate
calculations of theoretical line ratios for Fe~{\sc xiii}, with the most reliable
consisting of a 27 fine-structure level model ion, 
comprising the 3s$^{2}$3p$^{2}$, 3s3p$^{3}$ and 3s$^{2}$3p3d configurations,
with radiative rates being taken from the SUPERSTRUCTURE calculations of Young (2004),
and electron impact excitation
cross sections from the R-matrix results of Gupta \& Tayal (1998). However, Landi 
found that some of the Fe~{\sc xiii} line intensities showed large discrepancies between 
theory and observation, which could not be explained by blending, and was therefore
most likely due to 
errors in the adopted atomic data (see also Young et al. 1998 and Landi \&
Landini 1997).

More recently, Aggarwal \& Keenan (2004, 2005) have employed the relativistic 
GRASP and Dirac R-matrix codes to calculate radiative rates and electron impact 
excitation cross sections, respectively, for all transitions among the energetically
lowest 97 fine-structure levels of Fe\,{\sc xiii}. In this paper we use these data to
generate theoretical emission-line ratios for Fe\,{\sc xiii}, and compare these
with SERTS observations, to (i) investigate if previous discrepancies between theory and 
observation may be resolved, (ii) assess the importance of blending in the SERTS observations,
(iii) detect new Fe\,{\sc xiii} emission lines, and finally (iv) identify the best
Fe\,{\sc xiii} line pairs for use as electron density diagnostics.

\section[]{Observational data}

The first fully successful SERTS flight
was on 1989 May 5 
(Neupert et al. 1992; Thomas \& Neupert 1994), and carried a standard
gold-coated toroidal diffraction grating. It observed active region NOAA 5464 and detected hundreds of 
first-order emission
lines in the 235--450\,\AA\ wavelength range, as well as dozens of features spanning
170--225\,\AA, which appeared in second-order among the 340--450\,\AA\ first-order
lines. The spectrum was recorded
on Kodak 101--07 emulsion, at a spectral resolution of 50--80\,m\AA\ (FWHM) in first-order, and a spatial
resolution of approximately 7 arcsec (FWHM).

SERTS had additional technology-demonstration flights in 1991 
and 1993 (Brosius et al. 1996), carrying a multilayer-coated diffraction 
grating that enhanced the instrumental sensitivity over a limited range 
of its first-order bandpass.  However, over much of the range, these 
observations are actually at a somewhat reduced sensitivity compared to 
the 1989 data.  As a result, fewer emission lines were detected, and 
many of those that were measured had 
larger uncertainities (see, for example, the comparison of SERTS 1989, 1991 and 1993 
measurements in Keenan et al. 2003a,b and Keenan et al. 2005b). Consequently, the 1989 observations
(henceforth referred to as SERTS--89) provide the best SERTS spectrum for investigating emission
features in the first-order wavelength range 235--450\,\AA, and are hence employed in the present analysis.

During a rocket flight on 1995 May 15, SERTS observed active region NOAA 7870, once again 
recording the spectrum on Kodak 101--07 emulsion at a spatial
resolution of approximately 5 arcsec (Brosius, Davila \& Thomas 1998). However, this version of the
instrument incorporated a multilayer-coated toroidal diffraction grating that enhanced
its sensivitity to second-order features in the 170--225\,\AA\ wavelength range. This led to the 
detection of many second-order emission lines not seen on previous SERTS flights (Thomas \& Neupert 1994;
Brosius et al. 1996), and furthermore provided the highest spectral resolution (0.03\,\AA)
ever achieved for spatially resolved active region spectra in this wavelength range. The SERTS 1995
active region spectrum (henceforth SERTS--95) therefore provides the best observations for investigating
emission lines in the 170--225\,\AA\ region, and is employed in the present paper. Further details of the
observations, and the wavelength and absolute flux calibration procedures employed in the data reduction,
may be found in Brosius et al. (1998). Similar information for the SERTS--89 spectrum is available
from Thomas \& Neupert (1994). We note that the SERTS spectra have been calibrated using temperature
and density insensitive emission line ratios, including those from Fe\,{\sc xiii}.
As the relevant line ratio calculations may have 
changed since these calibrations were performed, and indeed several have (see, for example,
Keenan et al. 2005b), any comparison of theoretical and observed Fe\,{\sc xiii}
line ratios must be treated with some caution. However, even if the calibration were to change, this
would be unlikely to significantly affect line pairs which are close in wavelength. 

We have searched for Fe\,{\sc xiii} emission lines in the SERTS--89 and SERTS--95 spectra using the
identifications of Thomas \& Neupert (1994) and Brosius et al. (1998), as well as other sources including
the NIST database,\footnote{http:$//$physics.nist.gov$/$PhysRefData$/$} 
the latest version (V5.2) of the {\sc chianti}
database (Dere et al. 1997; Landi et al. 2006), and previous solar detections where available
(for example, those of Dere 1978). In Tables 1 and 2 we list the Fe\,{\sc xiii}
transitions found in the SERTS--95 and SERTS--89 spectra, respectively, along with the measured
wavelengths. We also indicate possible blending lines or alternative
identifications as suggested by Brosius et al. or Thomas \& Neupert in their original line lists 
for these observations. Note that we list two measured wavelengths for the same
3s$^{2}$3p$^{2}$ $^{3}$P$_{1}$--3s$^{2}$3p3d $^{3}$P$_{0}$ 
transition in the SERTS--95 spectrum, namely 202.43
and 203.17\,\AA. Brosius et al. identify Fe~{\sc xiii}
lines at 202.42 and 
203.16\,\AA\ in the SERTS--95 data, but Landi (2002) only lists the 203.16\,\AA\ emission feature 
(as does {\sc chianti}). We return to the identification of this line in Section 4.1.

\begin{table*}
 \centering
 \begin{minipage}{140mm}
  \caption{Fe\,{\sc xiii} line identifications in the second-order SERTS 1995 active region spectrum.}
  \begin{tabular}{cll}
  \hline
Wavelength (\AA)    &   Transition    & Note$^{a}$ 
\\
\hline
196.53 & 3s$^{2}$3p$^{2}$ $^{1}$D$_{2}$--3s$^{2}$3p3d $^{1}$F$_{3}$ 
\\
200.03 & 3s$^{2}$3p$^{2}$ $^{3}$P$_{1}$--3s$^{2}$3p3d $^{3}$D$_{2}$  
\\
201.13 & 3s$^{2}$3p$^{2}$ $^{3}$P$_{1}$--3s$^{2}$3p3d $^{3}$D$_{1}$ & 
Blended with Fe\,{\sc xii} line$^{b}$ 
\\
202.04 & 3s$^{2}$3p$^{2}$ $^{3}$P$_{0}$--3s$^{2}$3p3d $^{3}$P$_{1}$ 
\\
202.43 & 3s$^{2}$3p$^{2}$ $^{3}$P$_{1}$--3s$^{2}$3p3d $^{3}$P$_{0}$ 
\\
203.17 & 3s$^{2}$3p$^{2}$ $^{3}$P$_{1}$--3s$^{2}$3p3d $^{3}$P$_{0}$ & 
Identified as this transition by Landi (2002)
\\
203.79 & 3s$^{2}$3p$^{2}$ $^{3}$P$_{2}$--3s$^{2}$3p3d $^{3}$D$_{2}$ &
No line listed 
\\
203.83 & 3s$^{2}$3p$^{2}$ $^{3}$P$_{2}$--3s$^{2}$3p3d $^{3}$D$_{3}$ & 
Listed with 203.79\,\AA\ as a single feature
\\
204.26 & 3s$^{2}$3p$^{2}$ $^{3}$P$_{1}$--3s$^{2}$3p3d $^{1}$D$_{2}$ 
\\
204.95 & 3s$^{2}$3p$^{2}$ $^{3}$P$_{2}$--3s$^{2}$3p3d $^{3}$D$_{1}$  
\\
208.67 & 3s$^{2}$3p$^{2}$ $^{1}$S$_{0}$--3s$^{2}$3p3d $^{1}$P$_{1}$ & 
No line listed; identified as Ca\,{\sc xv} by Dere
(1978)
\\
209.63 & 3s$^{2}$3p$^{2}$ $^{3}$P$_{1}$--3s$^{2}$3p3d $^{3}$P$_{2}$  
\\
209.91 & 3s$^{2}$3p$^{2}$ $^{3}$P$_{2}$--3s$^{2}$3p3d $^{3}$P$_{1}$  
\\
213.77 & 3s$^{2}$3p$^{2}$ $^{3}$P$_{2}$--3s$^{2}$3p3d $^{3}$P$_{2}$ 
\\
216.90 & 3s$^{2}$3p$^{2}$ $^{1}$D$_{2}$--3s$^{2}$3p3d $^{3}$D$_{2,3}$ & Blended with Si\,{\sc viii} line 
\\
221.82 & 3s$^{2}$3p$^{2}$ $^{1}$D$_{2}$--3s$^{2}$3p3d $^{1}$D$_{2}$  
\\
\hline
\end{tabular}

$^{a}$From Brosius et al. (1998).
\\
$^{b}$From Thomas \& Neupert (1994). 
\end{minipage} 
\end{table*}

\begin{table*}
 \centering
 \begin{minipage}{140mm}
  \caption{Fe\,{\sc xiii} line identifications in the first-order SERTS 1989 active region spectrum.}
  \begin{tabular}{cll}
  \hline
Wavelength (\AA)    &   Transition    & Note$^{a}$ 
\\
\hline
240.72 & 3s$^{2}$3p$^{2}$ $^{3}$P$_{0}$--3s3p$^{3}$ $^{3}$S$_{1}$ 
\\
246.20 & 3s$^{2}$3p$^{2}$ $^{3}$P$_{1}$--3s3p$^{3}$ $^{3}$S$_{1}$ 
\\
251.94 & 3s$^{2}$3p$^{2}$ $^{3}$P$_{2}$--3s3p$^{3}$ $^{3}$S$_{1}$ 
\\
256.43 & 3s$^{2}$3p$^{2}$ $^{1}$D$_{2}$--3s3p$^{3}$ $^{1}$P$_{1}$ & Blended with Zn\,{\sc xx} line
\\
259.94 & 3s$^{2}$3p$^{2}$ $^{1}$D$_{2}$--3s$^{2}$3p3d $^{3}$F$_{2}$ & Unidentified line 
\\
272.21 & 3s$^{2}$3p$^{2}$ $^{1}$D$_{2}$--3s3p$^{3}$ $^{3}$S$_{1}$ & Unidentified line 
\\
288.56 & 3s$^{2}$3p$^{2}$ $^{1}$S$_{0}$--3s3p$^{3}$ $^{1}$P$_{1}$ & No line listed$^{b}$
\\
290.81 & 3s$^{2}$3p$^{2}$ $^{3}$P$_{2}$--3s3p$^{3}$ $^{1}$D$_{2}$ & No line listed 
\\
311.57 & 3s$^{2}$3p$^{2}$ $^{3}$P$_{1}$--3s3p$^{3}$ $^{3}$P$_{2}$ 
\\
312.17 & 3s$^{2}$3p$^{2}$ $^{3}$P$_{1}$--3s3p$^{3}$ $^{3}$P$_{1}$ 
\\
312.89 & 3s$^{2}$3p$^{2}$ $^{3}$P$_{1}$--3s3p$^{3}$ $^{3}$P$_{0}$ & Unidentified line$^{c}$ 
\\
318.12 & 3s$^{2}$3p$^{2}$ $^{1}$D$_{2}$--3s3p$^{3}$ $^{1}$D$_{2}$ 
\\
320.80 & 3s$^{2}$3p$^{2}$ $^{3}$P$_{2}$--3s3p$^{3}$ $^{3}$P$_{2}$ 
\\
321.46 & 3s$^{2}$3p$^{2}$ $^{3}$P$_{2}$--3s3p$^{3}$ $^{3}$P$_{1}$ 
\\
348.18 & 3s$^{2}$3p$^{2}$ $^{3}$P$_{0}$--3s3p$^{3}$ $^{3}$D$_{1}$ 
\\
359.64 & 3s$^{2}$3p$^{2}$ $^{3}$P$_{1}$--3s3p$^{3}$ $^{3}$D$_{2}$ 
\\
359.83 & 3s$^{2}$3p$^{2}$ $^{3}$P$_{1}$--3s3p$^{3}$ $^{3}$D$_{1}$ 
\\
368.16 & 3s$^{2}$3p$^{2}$ $^{3}$P$_{2}$--3s3p$^{3}$ $^{3}$D$_{3}$ & Blended with Cr\,{\sc xiii} line 
\\
413.00 & 3s$^{2}$3p$^{2}$ $^{1}$D$_{2}$--3s3p$^{3}$ $^{3}$D$_{3}$ &
\\
\hline
\end{tabular}

$^{a}$From Thomas \& Neupert (1994).
\\
$^{b}$Listed by Brosius et al. (1998) as an unidentified feature. 
\\
$^{c}$Subsequently identified by Brickhouse et al. (1995).
\end{minipage} 
\end{table*}

Intensities and observed
line widths (FWHM) of the Fe\,{\sc xiii} features are given in Tables 3 and
4 for the SERTS--95 and SERTS--89 datasets, respectively,
along with the associated 1$\sigma$ errors. These were measured
using modified versions of the Gaussian fitting routines employed by Thomas \& Neupert (1994).
Specifically, we have incorporated the ability to filter spectral noise at low 
intensities. By utilizing fitted Gaussian exponents around the line centre, it is 
possible to extrapolate wavelength and intensity values away from the 
emission line core, thus removing any artefacts associated with noise spikes 
and/or hot pixels which may affect the unfiltered spectral profile.
The intensities, FWHM values and their uncertainties listed in Tables 3 and
4 are hence somewhat different from those originally
reported in Thomas \& Neupert and Brosius et al. (1998). Also,
a uniform factor of 1.24 has been applied here to all
SERTS--89 intensities, reflecting a more recent re-evaluation of its 
absolute
calibration scale.
Even so, in all directly comparable cases, the resulting
line intensity values usually differ only slightly from those obtained using 
previously
published data.
The listed line widths include an instrumental broadening component due 
to the spectrometer optics, and so should be considered as only upper 
limits to the actual widths of the lines.

\begin{table}
 \centering
  \caption{Fe\,{\sc xiii} line intensities and widths from the SERTS 1995 active region spectrum.}
  \begin{tabular}{ccc}
  \hline
Wavelength    &   Intensity & Line width \\
(\AA) & (erg\,cm$^{-2}$\,s$^{-1}$\,sr$^{-1}$) & (m\AA) 
\\
\hline
196.53 & 134.5 $\pm$ 33.1 & 42 $\pm$ 7 
\\
200.03 & 305.0 $\pm$ 35.6 & 47 $\pm$ 4 
\\
201.13 & 469.4 $\pm$ 59.8 & 49 $\pm$ 3 
\\
202.04 & 1219.1 $\pm$ 135.6 & 49 $\pm$ 3 
\\
202.43 & 73.6 $\pm$ 17.1 & 38 $\pm$ 4 
\\
203.17 & 155.8 $\pm$ 23.9 & 37 $\pm$ 3 
\\
203.79 & 310.2 $\pm$ 43.9 & 46 $\pm$ 3 
\\
203.83 & 1547.2 $\pm$ 169.8 & 54 $\pm$ 4 
\\
204.26 & 192.0 $\pm$ 27.7 & 46 $\pm$ 5 
\\
204.95 & 268.7 $\pm$ 42.5 & 63 $\pm$ 6 
\\
208.67 & 44.6 $\pm$ 13.5 & 25 $\pm$ 5 
\\
209.63 & 210.1 $\pm$ 33.6 & 44 $\pm$ 4 
\\
209.91 & 236.7 $\pm$ 41.5 & 63 $\pm$ 7 
\\
213.77 & 93.8 $\pm$ 20.4 & 35 $\pm$ 4 
\\
216.90 & 54.3 $\pm$ 15.9 & 32 $\pm$ 4 
\\
221.82 & 219.4 $\pm$ 30.9 & 30 $\pm$ 3 
\\
\hline
\end{tabular}
\end{table}

\begin{table}
 \centering
  \caption{Fe\,{\sc xiii} line intensities and widths from the SERTS 1989 active region spectrum.}
  \begin{tabular}{ccc}
  \hline
Wavelength    &   Intensity & Line width \\
(\AA) & (erg\,cm$^{-2}$\,s$^{-1}$\,sr$^{-1}$) & (m\AA) 
\\
\hline
240.72 & 156.6 $\pm$ 60.3 & 74 $\pm$ 20 
\\
246.20 & 140.7 $\pm$ 41.7 & 42 $\pm$ 6 
\\
251.94 & 447.8 $\pm$ 66.3 & 84 $\pm$ 6 
\\
256.43 & 176.0 $\pm$ 63.4 & 96 $\pm$ 28 
\\
259.94 & 49.2 $\pm$ 19.8 & 30 $\pm$ 8 
\\
272.21 & 82.5 $\pm$ 19.2 & 50 $\pm$ 9 
\\
288.56 & 18.0 $\pm$ 8.2 & 45 $\pm$ 11 
\\
290.81 & 6.1 $\pm$ 3.2 & 25 $\pm$ 10 
\\
311.57 & 37.4 $\pm$ 11.8 & 66 $\pm$ 13
\\
312.17 & 104.7 $\pm$ 10.4 & 87 $\pm$ 5 
\\
312.89 & 57.7 $\pm$ 16.0 & 42 $\pm$ 8 
\\
318.12 & 108.5 $\pm$ 14.0 & 99 $\pm$ 8 
\\
320.80 & 212.8 $\pm$ 27.2 & 95 $\pm$ 6 
\\
321.46 & 41.8 $\pm$ 9.1 & 83 $\pm$ 10 
\\
348.18 & 158.2 $\pm$ 18.7 & 90 $\pm$ 5 
\\
359.64 & 182.9 $\pm$ 20.4 & 91 $\pm$ 5 
\\
359.83 & 27.9 $\pm$ 5.4 & 77 $\pm$ 8
\\
368.16 & 167.2 $\pm$ 29.9 & 83 $\pm$ 11
\\
413.00 & 8.8 $\pm$ 2.5 & 120 $\pm$ 15 
\\
\hline
\end{tabular}
\end{table}

We plot portions of the SERTS--89 and SERTS--95 spectra 
containing Fe\,{\sc xiii} features in Figs 1--5,
to show the quality of the observational data. In particular, we have been
able to identify both the 203.79 and 203.83\,\AA\ lines of Fe\,{\sc xiii} (Fig. 2), 
which previously have been listed as a single feature (see, for example, Brosius et al. 1998). 
We note that each SERTS spectrum exhibits
a background level due to film fog, scattered light and actual
solar continuum. The background was calculated using methods
detailed in Thomas \& Neupert (1994) and Brosius et al.,
and then subtracted from the initial spectrum, leaving
only an emission line spectrum (with noise) on a zero base level.
It is this zero base level which is shown in Figs 1--5.
We note that some of the measured Fe\,{\sc xiii} emission
lines, such as the 288.56\,\AA\ transition (Fig. 4), have
line intensities comparable to the noise fluctuations.
In these instances the reality of the line was confirmed
by a visual inspection of the original SERTS film.

\begin{figure}
\epsfig{file=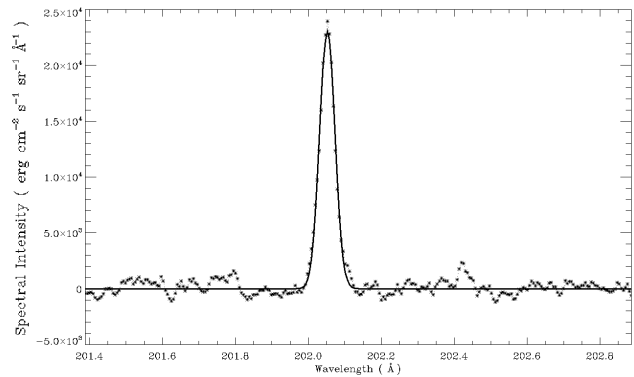,angle=0,width=8.5cm}
\caption{Plot of the SERTS 1995 active region spectrum in the
201.4--202.9\,\AA\ wavelength range.
The profile fit to the Fe\,{\sc xiii} 202.04\,\AA\ feature
is shown by a solid line.}
\end{figure}

\begin{figure}
\epsfig{file=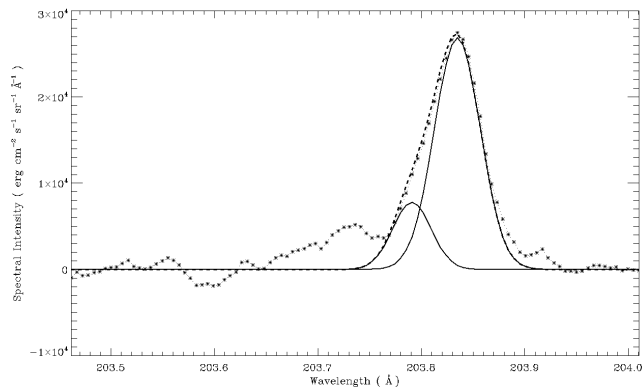,angle=0,width=8.5cm}
\caption{Plot of the SERTS 1995 active region spectrum in the
203.5--204.0\,\AA\ wavelength range.
The profile fit to the Fe\,{\sc xiii} 203.79 and
203.83\,\AA\ features
is shown by a dashed line, while the fits for the individual components
are shown by solid lines. Also clearly visible in the
figure is an unidentified feature at 203.73\,\AA.}
\end{figure}

\begin{figure}
\epsfig{file=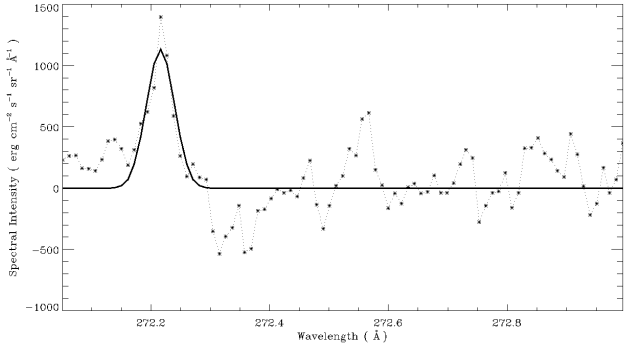,angle=0,width=8.5cm}
\caption{Plot of the SERTS 1989 active region spectrum in the
272.1--272.9\,\AA\ wavelength range.
The profile fit to the Fe\,{\sc xiii} 272.21\,\AA\ feature
is shown by a solid line.}
\end{figure}

\begin{figure}
\epsfig{file=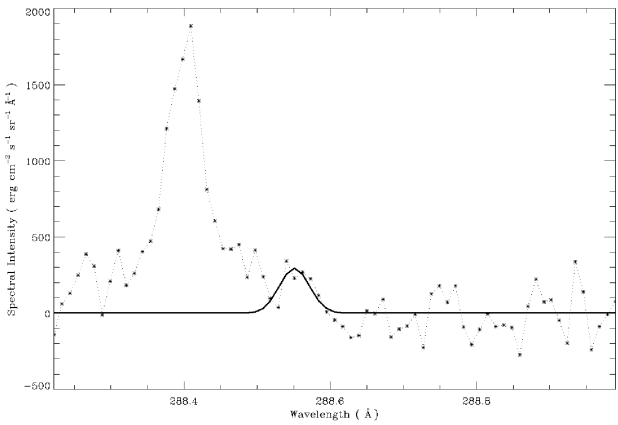,angle=0,width=8.5cm}
\caption{Plot of the SERTS 1989 active region spectrum in the
288.3--288.9\,\AA\ wavelength range.
The profile fit to the Fe\,{\sc xiii} 288.56\,\AA\ feature
is shown by a solid line. Also clearly visible in the
figure is the S\,{\sc xii} 288.40\,\AA\ line.}
\end{figure}

\begin{figure}
\epsfig{file=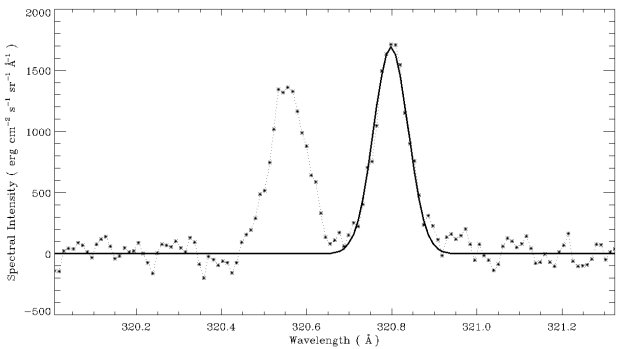,angle=0,width=8.5cm}
\caption{Plot of the SERTS 1989 active region spectrum in the
320.1--321.2\,\AA\ wavelength range.
The profile fit to the Fe\,{\sc xiii} 320.80\,\AA\ feature
is shown by a solid line. Also clearly visible in the
figure is the Ni\,{\sc xviii} 320.56\,\AA\ line.}
\end{figure}

\section{Theoretical line ratios}

The model ion for Fe~{\sc xiii} consisted of the 97
energetically lowest fine-structure levels, which are listed in Table 1 of
Aggarwal \& Keenan (2004), and include levels from the 
3s$^{2}$3p$^{2}$, 3s3p$^{3}$, 3s$^{2}$3p3d, 
3p$^{4}$, 3s3p$^{2}$3d and 3s$^{2}$3d$^{2}$ configurations.
Experimental energy levels, which are only available for a 
small number (22) of Fe\,{\sc xiii} states, were obtained from
the NIST database.
For the remaining values the theoretical results of Aggarwal \& Keenan (2004)
were adopted.

The electron impact excitation cross sections employed in the present paper
are the fully relativistic Dirac R-matrix code calculations of Aggarwal \& Keenan (2005).
For Einstein A-coefficients, Aggarwal \& Keenan (2004)
undertook two calculations with the fully relativistic {\sc grasp} code, namely one
considering the 97 fine-structure levels from the 6 configurations
listed above (termed GRASP6), and another
with 301 levels 
from the 13 configurations 
3s$^2$3p$^2$, 3s3p$^3$,
3s$^2$3p3d, 3p$^4$, 3s3p$^2$3d, 3s$^2$3d$^2$, 3p$^3$3d, 3s3p3d$^2$,
3s3d$^3$ and 3s$^2$3p4$\ell$ (GRASP13). Aggarwal \& Keenan noted that the
GRASP13 results represent a significant
improvement over the GRASP6 data, and they have therefore been adopted
in the present paper. 
Unfortunately however, Aggarwal \& Keenan only published their GRASP6 results,
and hence in Tables 5 and 6 we provide the GRASP13 calculations for all 4656 
transitions among the 97 levels considered in our model ion. Complete versions of these
tables are available in the electronic version of the paper, with only
sample results presented in the hardcopy edition. The indices used to 
represent the lower and upper levels of a transition 
have been defined in Table 1 of Aggarwal \& Keenan. We note that radiative data 
for all 45150 transitions 
among the 301 levels considered in our GRASP13 calculations are available
in electronic form on request from one of the authors (K.Aggarwal@qub.ac.uk).

\begin{table*}
 \centering
  \caption{Transition energies/wavelengths ($\lambda_{ij}$ in $\AA$), radiative rates (A$_{ji}$ in s$^{-1}$), 
oscillator strengths (f$_{ij}$, dimensionless), and     
line strengths (S, in atomic units) for a sample of 
electric dipole (E1) and magnetic quadrupole (M2) transitions in Fe\,{\sc xiii}. ($a{\pm}b \equiv a{\times}$10$^{{\pm}b}$).
}
\begin{tabular}{rrrrrrrrr}                                                                                                                                                                                                                                                                                        
\hline                                                                                                                                                          
$i$ & $j$ & $\lambda_{ij}$ & A$^{{\rm E1}}_{ji}$  & f$^{{\rm E1}}_{ij}$ & S$^{{\rm E1}}$ & A$^{{\rm M2}}_{ji}$  & f$^{{\rm M2}}_{ij}$ & S$^{{\rm M2}}$ \\                 
\hline                                                                                                                                                   
    1 &    6 &  4.802$+$02 &  0.000$+$00 &  0.000$+$00 &  0.000$+$00 &  5.771$-$01 &  9.975$-$11 &  4.941$+$00
 \\       
    1 &    7 &  3.505$+$02 &  1.224$+$09 &  6.763$-$02 &  7.804$-$02 &  0.000$+$00 &  0.000$+$00 &  0.000$+$00
 \\       
    1 &    8 &  3.503$+$02 &  0.000$+$00 &  0.000$+$00 &  0.000$+$00 &  3.193$+$00 &  2.937$-$10 &  5.647$+$00
 \\       
    1 &   11 &  3.040$+$02 &  1.262$+$09 &  5.244$-$02 &  5.248$-$02 &  0.000$+$00 &  0.000$+$00 &  0.000$+$00
 \\       
    1 &   12 &  3.034$+$02 &  0.000$+$00 &  0.000$+$00 &  0.000$+$00 &  1.292$+$00 &  8.919$-$11 &  1.115$+$00 
\\       
    1 &   13 &  2.754$+$02 &  0.000$+$00 &  0.000$+$00 &  0.000$+$00 &  4.536$+$00 &  2.579$-$10 &  2.410$+$00
 \\       
    1 &   14 &  2.306$+$02 &  0.000$+$00 &  0.000$+$00 &  0.000$+$00 &  5.677$+$00 &  2.263$-$10 &  1.241$+$00
 \\       
    1 &   15 &  2.367$+$02 &  7.507$+$09 &  1.892$-$01 &  1.475$-$01 &  0.000$+$00 &  0.000$+$00 &  0.000$+$00
 \\       
    1 &   18 &  2.241$+$02 &  8.513$+$08 &  1.923$-$02 &  1.419$-$02 &  0.000$+$00 &  0.000$+$00 &  0.000$+$00
 \\       
    1 &   19 &  2.023$+$02 &  0.000$+$00 &  0.000$+$00 &  0.000$+$00 &  9.328$-$01 &  2.860$-$11 &  1.059$-$01
 \\       
    1 &   20 &  1.990$+$02 &  4.549$+$10 &  8.100$-$01 &  5.306$-$01 &  0.000$+$00 &  0.000$+$00 &  0.000$+$00
 \\       
    1 &   22 &  1.971$+$02 &  0.000$+$00 &  0.000$+$00 &  0.000$+$00 &  7.364$-$03 &  2.143$-$13 &  7.336$-$04 
\\       
    1 &   23 &  1.942$+$02 &  1.245$+$10 &  2.112$-$01 &  1.350$-$01 &  0.000$+$00 &  0.000$+$00 &  0.000$+$00 
\\       
    1 &   24 &  1.931$+$02 &  0.000$+$00 &  0.000$+$00 &  0.000$+$00 &  7.913$+$00 &  2.211$-$10 &  7.116$-$01 
\\       
    1 &   28 &  1.719$+$02 &  4.218$+$08 &  5.603$-$03 &  3.170$-$03 &  0.000$+$00 &  0.000$+$00 &  0.000$+$00 
\\       

\hline
\end{tabular}
\end{table*}

\begin{table*}
 \centering
  \caption{Transition energies/wavelengths ($\lambda_{ij}$ in $\AA$), radiative rates (A$_{ji}$ in s$^{-1}$)),
 oscillator strengths (f$_{ij}$, dimensionless), and     
line strengths (S, in atomic units) for a sample of 
electric quadrupole (E2) and magnetic dipole (M1) transitions in Fe\,{\sc xiii}. ($a{\pm}b \equiv a{\times}$10$^{{\pm}b}$).}

\begin{tabular}{rrrrrrrrr}                                                                                                                                                                                                                                                                                            
\hline                                                                                                                                                          
$i$ & $j$ & $\lambda_{ij}$ & A$^{{\rm E2}}_{ji}$  & f$^{{\rm E2}}_{ij}$ & S$^{{\rm E2}}$ & A$^{{\rm M1}}_{ji}$  & f$^{{\rm M1}}_{ij}$ & S$^{{\rm M1}}$ \\                 
\hline  
    1 &    2 &  1.092$+$04 &  0.000$+$00 &  0.000$+$00 &  0.000$+$00 &  1.332$+$01 &  7.147$-$07 &  1.930$+$00
\\       
    1 &    3 &  5.382$+$03 &  2.572$-$03 &  5.584$-$11 &  5.183$-$02 &  0.000$+$00 &  0.000$+$00 &  0.000$+$00
\\       
    1 &    4 &  1.994$+$03 &  2.514$-$01 &  7.492$-$10 &  3.537$-$02 &  0.000$+$00 &  0.000$+$00 &  0.000$+$00
 \\       
    1 &   27 &  1.644$+$02 &  9.052$+$04 &  1.833$-$06 &  4.848$-$02 &  0.000$+$00 &  0.000$+$00 &  0.000$+$00
 \\       
    1 &   29 &  1.612$+$02 &  0.000$+$00 &  0.000$+$00 &  0.000$+$00 &  8.133$+$00 &  9.500$-$11 &  3.786$-$06
 \\       
    1 &   31 &  1.582$+$02 &  0.000$+$00 &  0.000$+$00 &  0.000$+$00 &  1.470$-$01 &  1.655$-$12 &  6.476$-$08
 \\       
    1 &   32 &  1.577$+$02 &  6.916$+$02 &  1.289$-$08 &  3.010$-$04 &  0.000$+$00 &  0.000$+$00 &  0.000$+$00
 \\       
    1 &   33 &  1.576$+$02 &  4.851$+$03 &  9.029$-$08 &  2.104$-$03 &  0.000$+$00 &  0.000$+$00 &  0.000$+$00
 \\       
    1 &   38 &  1.521$+$02 &  0.000$+$00 &  0.000$+$00 &  0.000$+$00 &  1.839$-$02 &  1.912$-$13 &  7.191$-$09
 \\       
    1 &   39 &  1.518$+$02 &  8.246$+$02 &  1.425$-$08 &  2.969$-$04 &  0.000$+$00 &  0.000$+$00 &  0.000$+$00
 \\       
    1 &   42 &  1.456$+$02 &  2.378$+$05 &  3.778$-$06 &  6.941$-$02 &  0.000$+$00 &  0.000$+$00 &  0.000$+$00 
\\       
    1 &   46 &  1.410$+$02 &  1.708$+$01 &  2.545$-$10 &  4.251$-$06 &  0.000$+$00 &  0.000$+$00 &  0.000$+$00 
\\       
    1 &   47 &  1.406$+$02 &  0.000$+$00 &  0.000$+$00 &  0.000$+$00 &  8.572$+$00 &  7.620$-$11 &  2.649$-$06
 \\       
    1 &   49 &  1.347$+$02 &  9.169$+$04 &  1.248$-$06 &  1.817$-$02 &  0.000$+$00 &  0.000$+$00 &  0.000$+$00
 \\       
    1 &   50 &  1.331$+$02 &  0.000$+$00 &  0.000$+$00 &  0.000$+$00 &  2.064$-$01 &  1.644$-$12 &  5.412$-$08
 \\

\hline
\end{tabular}
\end{table*}

Proton
impact excitation is only important for the fine-structure transitions 
among the ground-state 3s$^{2}$3p$^{2}$ $^{3}$P$_{0,1,2}$ levels. In the 
present paper we have employed the quantal calculations of Faucher (1975)
as given in Faucher \& Landman (1977).

Using the above atomic data, in conjunction
with a recently updated version of the statistical equilibrium
code of Dufton (1977), relative Fe\,{\sc xiii}
level populations and hence emission-line strengths were calculated
for a grid of 
electron temperature (T$_{e}$) and density (N$_{e}$) values, with
T$_{e}$ = 10$^{6.0}$, 10$^{6.2}$ and 10$^{6.4}$\,K, 
and N$_{e}$ = 10$^{8}$--10$^{13}$\,cm$^{-3}$
in steps of 0.1 dex.
The adopted temperature range covers that over which Fe\,{\sc xiii}
has a fractional abundance in ionization equilibrium of 
N(Fe\,{\sc xiii})/N(Fe) $\geq$ 0.025 (Mazzotta et al. 1998), and hence should be
appropriate to most coronal-type plasmas.
Our results are far too extensive to reproduce here, as with 97 fine-structure levels in our 
calculations we have intensities for 4656 transitions at each of the 
153 possible (T$_{e}$, N$_{e}$) combinations considered. 
However, results 
involving any line pair, in either photon or energy units, are freely available from
one of the authors (FPK) by email on request.
In addition, we note that we have Fe\,{\sc xiii} atomic data for the electron
temperature range T$_{e}$ = 10$^{5.9}$--10$^{6.7}$~K in steps of 0.1 dex, and hence can quickly generate
relative line strengths at additional temperatures, if requested.

In Figs 6--8 we plot some sample theoretical 
emission line ratios as a function of both T$_{e}$ and N$_{e}$, to illustrate their
sensitivity to the adopted
plasma parameters. The transitions corresponding to the wavelengths
given in the figures are listed in Tables 1 or 2.
Given expected errors in the adopted atomic data of typically
$\pm$20 per cent (see the references above), we would expect the
theoretical ratios to be accurate to better than $\pm$30 per cent.
An inspection of Figs 6--8 reveals that 203.83/202.04 is potentially a particularly good
N$_{e}$--diagnostic, involving lines which are very close in wavelength, 
varying by a factor of 
43 between N$_{e}$ = 10$^{8}$ and 10$^{11}$\,cm$^{-3}$, and
showing relatively little sensitivity to the adopted temperature.
For example, changing T$_{e}$ from 10$^{6.2}$ to 10$^{6.4}$\,K (i.e. by 60 per cent)
would lead to a less than 0.1 dex difference in the derived
value of N$_{e}$ over the density interval
N$_{e}$ = 10$^{9}$--10$^{11}$\,cm$^{-3}$. We discuss the
usefulness of the Fe\,{\sc xiii} transitions as density diagnostics in more detail
in Section 4.3.

\begin{figure}
\epsfig{file=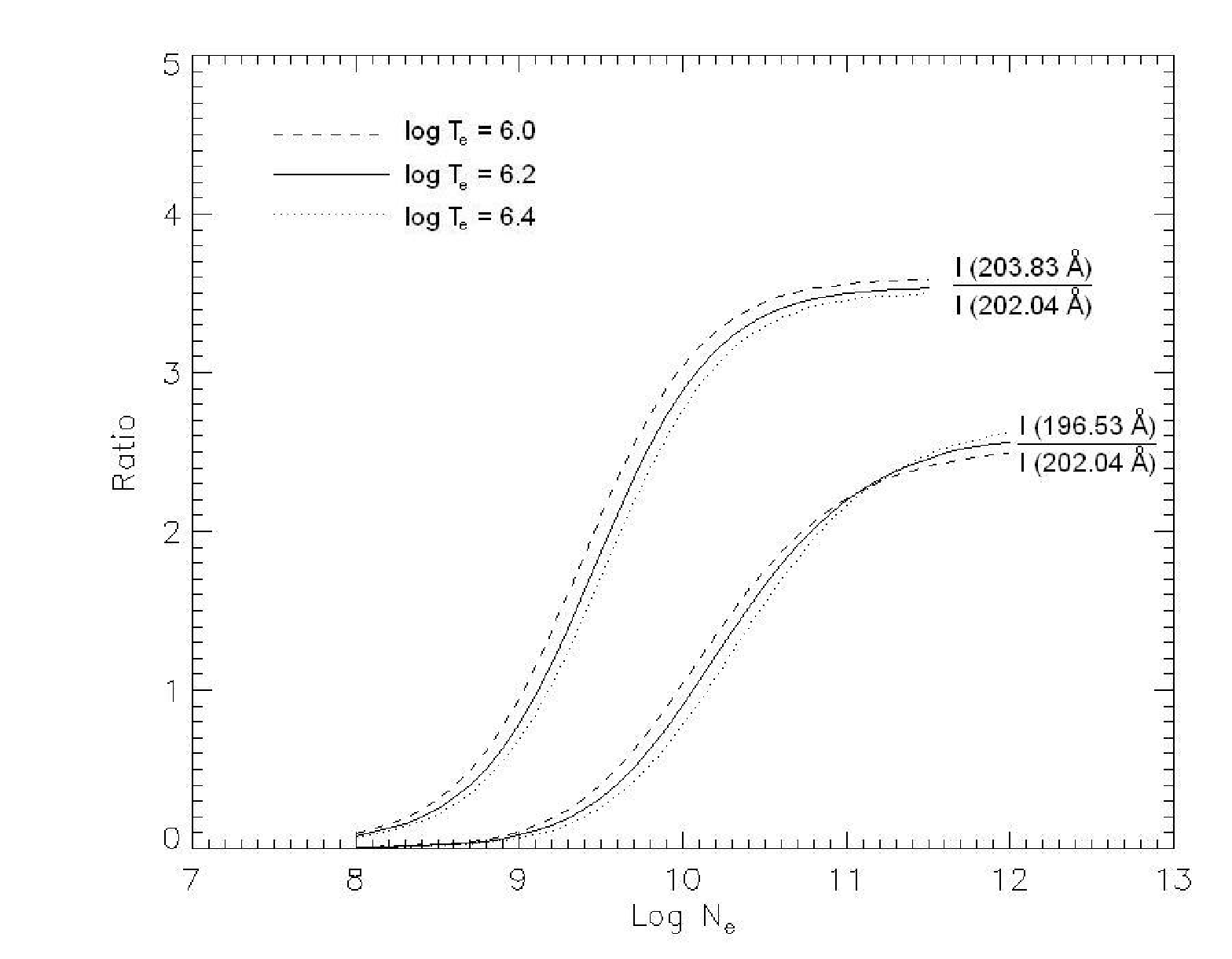,angle=0,width=8.5cm}
\caption{The theoretical Fe\,{\sc xiii}
emission line intensity ratios
I(203.83\,\AA)/I(202.04\,\AA) and I(196.53\,\AA)/I(202.04\,\AA),
where I is in energy units,
plotted as a function of logarithmic electron density
(N$_{e}$ in cm$^{-3}$) at the temperature
of maximum Fe\,{\sc xiii} fractional abundance in ionization
equilibrium, T$_{e}$ = 10$^{6.2}$\,K (Mazzotta et al.
1998), plus $\pm$0.2 dex about this
value. }
\end{figure}

\begin{figure}
\epsfig{file=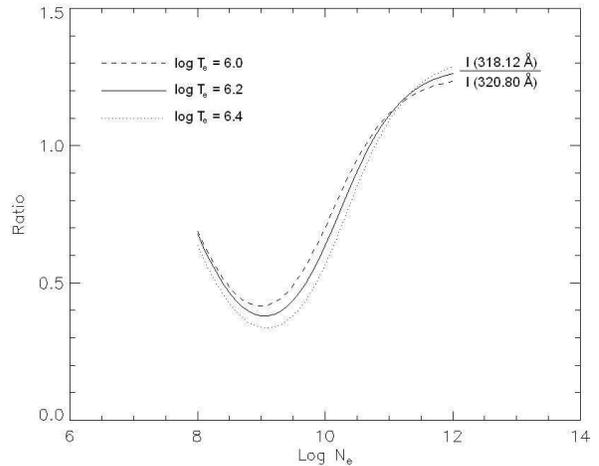,angle=0,width=8.5cm}
\caption{Same as Fig. 6 except for the 
I(318.12\,\AA)/I(320.80\,\AA) intensity ratio.}
\end{figure}

\begin{figure}
\epsfig{file=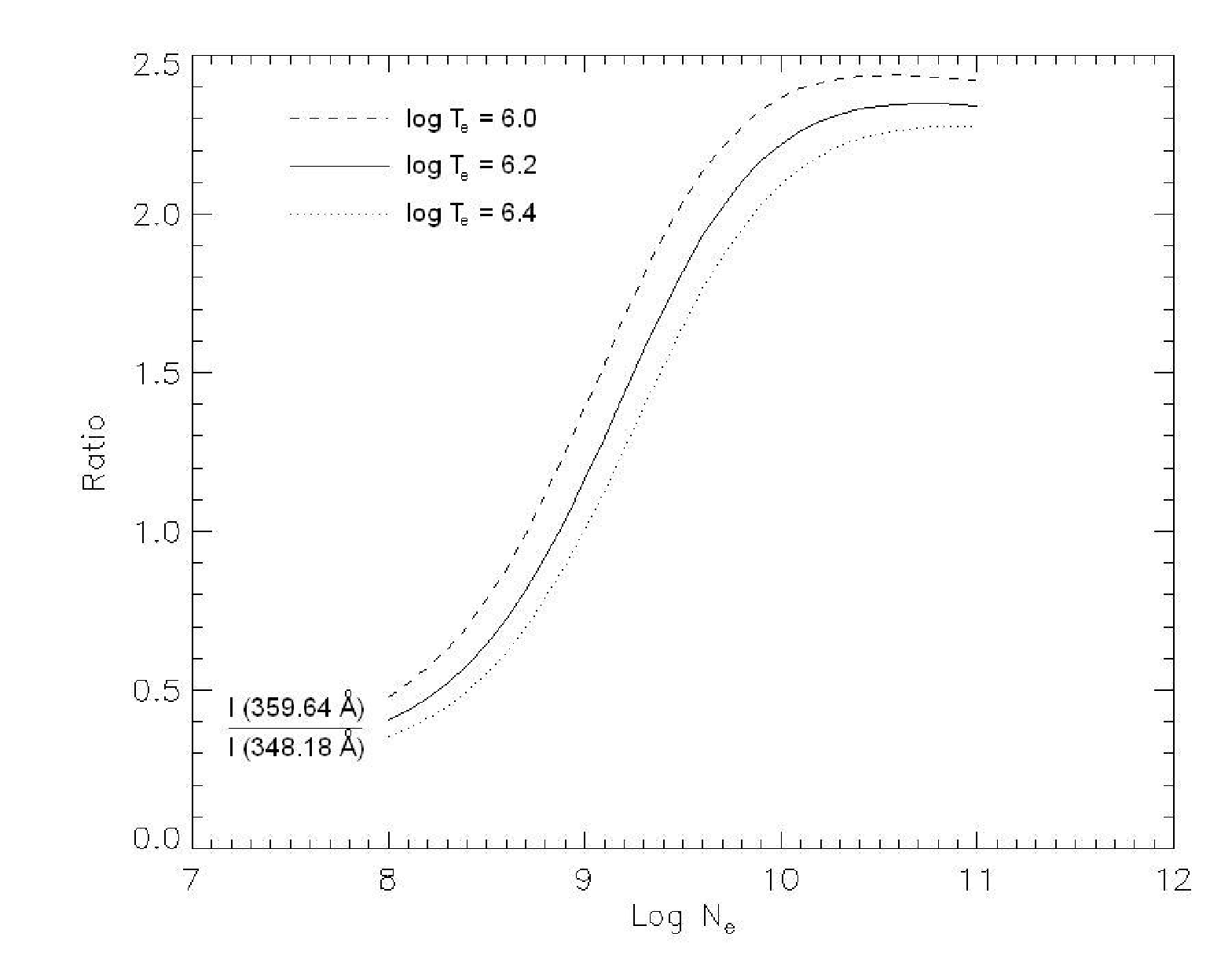,angle=0,width=8.5cm}
\caption{Same as Fig. 6 except for the 
I(359.64\,\AA)/I(348.18\,\AA) intensity ratio.}
\end{figure}

\section{Results and discussion}

Emission-line ratios may be categorised as 3 types, namely:

\begin{enumerate}

\item branching ratios, i.e. those which are predicted to be constant as the relevant transitions
arise from common upper levels;
\item
those which are predicted to be only weakly sensitive to the adopted electron temperature
and density over the range of plasma parameters of interest;
\item
those which are predicted to be strongly N$_{e}$--sensitive, and hence potentially
provide useful diagnostics.

\end{enumerate}

For a comparison of theory and observation, in order to identify blends and/or problems with
atomic data, clearly the line ratios which fall into categories 1 and 2 are the most useful,
as one does not need to reliably know the plasma electron temperature and density 
in order to calculate the line ratio.
Accordingly, in Tables 7 and 8 we list a range of observed line ratios for the SERTS--95 and SERTS--89
active regions, respectively (along with the associated 1$\sigma$ errors), which fulfill the criteria for 
category 1 or 2. Also listed in the tables 
are the predicted theoretical ratios both from the present calculations
and the latest version (V5.2) of the {\sc chianti} database (Dere et al. 1997; Landi et al. 2006),
which employs the 27-level model ion and atomic data discussed by Landi (2002). 
For category 2 ratios, we have defined `weakly sensitive' as being those which
are predicted to vary by less than $\pm$20 per cent when the electron density 
is changed by a factor of 2 (i.e. $\pm$0.3 dex). Electron density estimates for the SERTS--89 and
SERTS--95 active regions from line ratios in species formed at similar temperatures to Fe~{\sc xiii}
vary widely, with log N$_{e}$ $\simeq$ 8.8--9.8 for 
SERTS--89 
(Young et al. 1998) and log N$_{e}$ $\simeq$ 9.0--9.7 for SERTS--95 (Brosius et al. 1998).
However, most of the derived densities are consistent with log N$_{e}$ = 9.4$\pm$0.3 for both active regions.
The theoretical results in Tables 7 and 8 have therefore been calculated at the temperature of
maximum Fe\,{\sc xiii} fractional abundance in ionization equilibrium, 
T$_{e}$ = 10$^{6.2}$\,K (Mazzotta et al. 1998), and for N$_{e}$ = 10$^{9.4}$\,cm$^{-3}$. However we note that
changing the adopted temperature by $\pm$0.2 dex or the density by $\pm$0.5 dex would not significantly alter 
the discussions below.

\begin{table}
 \centering
\begin{minipage}{80mm}
  \caption{Comparison of theory and observation for emission-line intensity
ratios from the SERTS 1995 active region 
spectrum.}
  \begin{tabular}{cccc}
  \hline
Line ratio &   Observed & Present & {\sc chianti} 
\\
& & theory & theory
\\
\hline
\multicolumn{4}{l}{(i) Ratios of lines with common upper levels:}
\\[5pt]
200.03/203.79 & 0.98 $\pm$ 0.18 & 0.84 & 0.73
\\
201.13/204.95 & 1.7 $\pm$ 0.3 & 3.7 & 3.3 
\\
202.04/209.91 & 5.2 $\pm$ 1.1 & 5.1 & 6.7
\\
204.26/221.82 & 0.88 $\pm$ 0.18 & 0.38 & 0.59
\\
209.63/213.77 & 2.2 $\pm$ 0.6 & 1.1 & 1.0
\\[10pt]
\multicolumn{4}{l}{(ii) Ratios which are only weakly N$_{e}$--dependent:$^{a}$}
\\[5pt]
200.03/201.13 & 0.65 $\pm$ 0.11 & 0.74 & 0.75
\\
202.43/200.03 & 0.24 $\pm$ 0.06 & 0.49 & 0.53 
\\
203.17/200.03 & 0.51 $\pm$ 0.10 & 0.49 & 0.53 
\\
203.79/203.83 & 0.20 $\pm$ 0.04 & 0.32 & 0.33
\\
204.26/200.03 & 0.63 $\pm$ 0.12 & 0.32 & 0.63 
\\
204.95/200.03 & 0.88 $\pm$ 0.17 & 0.37 & 0.40
\\
208.67/200.03 & 0.15 $\pm$ 0.05 & 0.053 & 0.028 
\\
209.63/204.26 & 1.1 $\pm$ 0.2 & 2.9 & 1.2
\\
209.63/221.82 & 0.96 $\pm$ 0.20 & 1.1 & 0.72
\\
216.90/203.83 & 0.035 $\pm$ 0.011 & 0.074 & 0.10
\\
216.90/213.77 & 0.58 $\pm$ 0.21 & 0.33 & 0.57
\\
\hline
\end{tabular}

$^{a}$Present theoretical ratios and those from {\sc chianti} calculated at T$_{e}$ = 10$^{6.2}$\,K and
N$_{e}$ = 10$^{9.4}$\,cm$^{-3}$.
\end{minipage} 
\end{table}

\begin{table}
 \centering
\begin{minipage}{80mm}
  \caption{Comparison of theory and observation for emission-line intensity
ratios from the SERTS 1989 active region 
spectrum.}
  \begin{tabular}{cccc}
  \hline
Line ratio &   Observed & Present & {\sc chianti} 
\\
& & theory & theory
\\
\hline
\multicolumn{4}{l}{(i) Ratios of lines with common upper levels:}
\\[5pt]
240.72/246.20 & 1.1 $\pm$ 0.5 & 0.41 & 0.42
\\
246.20/251.94 & 0.31 $\pm$ 0.10 & 0.53 & 0.52 
\\
256.43/288.56 & 9.8 $\pm$ 5.7 & 7.6 & 7.8
\\
272.21/251.94 & 0.18 $\pm$ 0.05 & 0.018 & 0.018
\\
290.81/318.12 & 0.056 $\pm$ 0.030 & 0.020 & 0.022
\\
311.57/320.80 & 0.18 $\pm$ 0.06 & 0.14 & 0.13 
\\
321.46/312.17 & 0.40 $\pm$ 0.10 & 0.50 & 0.46
\\
359.83/348.18 & 0.18 $\pm$ 0.04 & 0.25 & 0.25
\\
413.00/368.16 & 0.053 $\pm$ 0.018 & 0.056 & 0.068 
\\[10pt]
\multicolumn{4}{l}{(ii) Ratios which are only weakly N$_{e}$--dependent:$^{a}$}
\\[5pt]
259.94/251.94 & 0.11 $\pm$ 0.05 & 0.10 & 0.016
\\
272.21/256.43 & 0.47 $\pm$ 0.20 & 0.020 & 0.098
\\
272.21/359.64 & 0.45 $\pm$ 0.12 & 0.025 & 0.11
\\
288.56/359.64 & 0.098 $\pm$ 0.046 & 0.16 & 0.11
\\
312.89/320.80 & 0.27 $\pm$ 0.08 & 0.24 & 0.28
\\
318.12/368.16 & 0.65 $\pm$ 0.14 & 0.33 & 0.29
\\
321.46/348.18 & 0.26 $\pm$ 0.06 & 0.30 & 0.31
\\
359.83/312.17 & 0.27 $\pm$ 0.06 & 0.42 & 0.36
\\
359.83/359.64 & 0.15 $\pm$ 0.03 & 0.15 & 0.27
\\
368.16/359.64 & 0.91 $\pm$ 0.19 & 1.2 & 0.95
\\
\hline
\end{tabular}

$^{a}$Present theoretical ratios and those from {\sc chianti} calculated at T$_{e}$ = 10$^{6.2}$\,K and
N$_{e}$ = 10$^{9.4}$\,cm$^{-3}$.
\end{minipage} 
\end{table}

\subsection{SERTS--95 active region spectrum: transitions in the 170--225\,\AA\ wavelength range} 

In the SERTS--95 second-order spectral region, only 
3s$^{2}$3p$^{2}$--3s$^{2}$3p3d transitions of Fe~{\sc xiii} are detected, spanning the
wavelength interval 196--222\,\AA\ (see Table 1). 
A comparison of the observed and theoretical 201.13/204.95 
ratios from Table 7 indicate that the measured 204.95\,\AA\ line intensity is too strong, supported 
by the fact that the observed 204.95/200.03 ratio is much larger than the theoretical value. Also, 
the 200.03/201.13 ratio measurement is in good agreement with the present calculations and those from
{\sc chianti}, indicating that both the 200.03 and 201.13\,\AA\ lines are reliably detected and modelled.
To identify possible blending species and assess their impact, we have generated
a synthetic active region spectrum using {\sc chianti}. However, no significant emission features are
predicted by {\sc chianti} close to 204.95\,\AA, apart from the Fe\,{\sc xiii} line. Young et al. (1998)
have suggested that the blending could be due to an unidentified first-order line at 409.90\,\AA, which
would appear in second-order at 204.95\,\AA. Although {\sc chianti}
has no suitable blending species at this wavelength, the database of 
van Hoof\footnote{http:$//$www.pa.uky.edu$/$$\sim$peter$/$atomic$/$} lists the Fe\,{\sc v}
 3d$^{4}$ $^{3}$F$_{2}$--3d$^{3}$4p $^{3}$G$_{3}$ line, which could be a possibility.
However, to properly assess this will need to await the calculation of accurate atomic data for
Fe\,{\sc v} and the inclusion of these in {\sc chianti}. We also note that although Thomas \& Neupert
(1994) list the 201.13\,\AA\ line as being blended with an Fe\,{\sc xii}
transition, the {\sc chianti} synthetic spectrum indicates that Fe\,{\sc xii} is responsible for
less than 3 per cent of the total line intensity, which is supported by the good agreement
found here between theory and observation. 

The good agreement between theory and observation for the 200.03/203.79 ratio
confirms our identification of the 203.79\,\AA\ transition, the first time (to our knowledge)
that this line has been 
detected separately from the 203.83\,\AA\ feature.
Similarly, there is no discrepancy between theory and observation for the 202.04/209.91 ratio,
indicating that both lines are reliably detected and free from problems. However, the experimental
value of the 204.26/221.82 ratio is much larger than theory, with the discrepancy with the present calculations
being larger than that with {\sc chianti}. 
This could be due to blending, as the measured 204.26/200.03 ratio is also larger than
theory, 
although in this instance there is good agreement with the {\sc chianti} prediction.
By contrast, the 209.63/221.82 experimental ratio shows a slightly smaller discrepancy 
with the present calculations, while the 209.63/204.26 
measurement is lower than the present calculations and in better agreement with {\sc chianti}.
However, the measured 209.63/213.77 ratio disagrees with both the present calculations and those from
{\sc chianti}. Landi (2002) note problems with the 204.26, 209.63 and 221.82\,\AA\ 
lines, which cannot be due to blending and hence are still clearly unresolved 
by the present analysis.

The good agreement of the observed 203.17/200.03 ratio with theory, as opposed to the discrepancy 
found for 202.43/200.03, indicates that the 203.17\,\AA\ feature is the Fe\,{\sc xiii}
line.
Although {\sc chianti} does not list a possible identification for the 202.43~\AA\ line, 
the list of van Hoof notes the presence of the Fe\,{\sc xv}
3s3d $^{3}$D$_{3}$--3p3d $^{3}$F$_{2}$ transition at 404.84\,\AA, which could be the 202.43~\AA\
feature in first-order. 
However, Keenan et al. (2005b) predicts the 404.84\,\AA\ line intensity to be only
2$\times$10$^{-5}$ that of Fe\,{\sc xv} 417.25\,\AA, which has I = 420
erg\,cm$^{-2}$\,s$^{-1}$\,sr$^{-1}$ (Thomas \& Neupert 1994). Hence Fe\,{\sc xv} is
far too weak to account
for the observed feature at 202.43~\AA.
Interestingly however, 
we note that there is another Fe\,{\sc v}
transition in the database of van Hoof predicted to lie at 404.87~\AA,  
namely 3d$^{4}$ $^{3}$D$_{2}$--3d$^{3}$4p $^{3}$D$_{1}$,
which will appear in second-order at 202.44\,\AA.

The experimental 208.67/200.03 ratio is larger than theory, but our results indicate that
Fe\,{\sc xiii}
contributes as much as 50 per cent to the 208.67\,\AA\ intensity, and the line is not due solely to
Ca\,{\sc xv}.
This is in agreement with the {\sc chianti}
synthetic spectrum, which indicates that Fe~{\sc xiii} should be responsible for around 80 per cent
of the total 208.67\,\AA\ intensity in an active region. 
However, we note that Ca\,{\sc xv} is known to dominate this emission line
in solar flares (Keenan et al. 2003c).

The observed and theoretical 216.90/213.77 ratios are in reasonable agreement, indicating that
both lines are reliably detected and free from blends. This is in contrast to Brosius et al. (1998),
who list 216.90\,\AA\ as a blend with Si\,{\sc viii}.
However, given the large observational error for the 216.90/213.77 ratio, we note that
Si\,{\sc viii}
could make as much as a 25 per cent contribution to the 216.90\,\AA\ 
intensity. In fact, {\sc chianti}
synthetic spectra indicate that it may make up to a 40 per cent contribution to this line
in the emission from some solar plasmas. 

Both the 216.90/203.83 and 203.79/203.83 observed ratios are smaller than predicted by theory, 
suggesting that the 203.83\,\AA\ transition may be blended, although this seems unlikely
for such a strong line. The {\sc chianti} synthetic spectrum indicates no suitable blending species
in either first- or second-order. However, once again the line list of van Hoof notes an
Fe\,{\sc v} transition at 407.65\,\AA, namely 
3d$^{4}$ $^{3}$D$_{1}$--3d$^{3}$4p $^{3}$P$_{0}$, which will 
appear in second-order at 203.83\,\AA.
 Although it must be considered improbable that Fe~{\sc v} 
is responsible for this and the other possible blends discussed above, 
these can only be ruled out when the ion is included in {\sc chianti}.

Unfortunately, the 196.53\,\AA\ line is predicted to be strongly N$_{e}$--sensitive 
when ratioed against any other transition of Fe\,{\sc xiii}
detected in the SERTS--95 spectrum. Hence 
it is not possible to generate 
ratios which are predicted to be independent (or nearly independent) of the adopted
density.
However, electron density diagnostics generated using the 196.53\,\AA\ feature yield 
results consistent
with other ratios (see Section 4.3), and we therefore 
believe the line is free of problems and 
significant blending. We note that the {\sc chianti} synthetic spectrum indicates no likely blending
species in either first- or second-order, and neither do other line lists. 
 
From the above, the Fe\,{\sc xiii} transitions which appear to be free from problems and significant
blending are: 196.53, 200.03, 201.13, 202.04, 203.17, 203.79, 209.91, 213.77\,\AA\ and
216.90\,\AA.
However, 
there are inconsistencies with the following lines: 203.83, 204.26, 204.95, 208.67, 209.63
and 221.82\,\AA. 
In the case of the 208.67\,\AA\ line, the problem is clearly due to a known blend, while for
203.83 and 204.95\,\AA, there may be unidentified blends present.
This leaves the 204.26, 209.63 and 221.82\,\AA\ transitions, where the problems cannot be due to blending
and likely arise from errors in the atomic data, as also concluded
by Landi (2002). It is interesting to note that two of these transitions (204.26 and
221.82\,\AA) have 3s$^{2}$3p3d $^{1}$D$_{2}$
as their upper level, and we would suggest that further atomic physics calculations (especially for
A-values) should pay particular attention to these. 

\subsection{SERTS--89 active region spectrum: transitions in the 235--450\,\AA\ wavelength range} 

In the 235--450\,\AA\ wavelength region covered by SERTS--89 in first-order, the
3s$^{2}$3p$^{2}$--3s3p$^{3}$ transitions of Fe\,{\sc xiii} are primarily detected, 
although we also have the provisional identification of one 
3s$^{2}$3p$^{2}$--3s$^{2}$3p3d line. All of the detected Fe\,{\sc xiii}
features lie in the 240--413\,\AA\ range (see Table 2). A comparison of the observed and theoretical
240.72/246.20 ratios in Table 8 reveals that the former is too large. This is probably due to 
blending of the 240.72\,\AA\ feature, as the 246.20/251.94 ratio shows reasonable agreement 
between theory and observation, indicating no problems with the 246.20 and 251.94\,\AA\ transitions.
The {\sc chianti} synthetic active region spectrum predicts that the Fe\,{\sc xii}
3s$^{2}$3p$^{3}$ $^{2}$D$_{5/2}$--3s$^{2}$3p$^{2}$3d $^{2}$F$_{7/2}$ transition 
makes a 25 per cent contribution to the Fe\,{\sc xii}/{\sc xiii} 240.72\,\AA\ intensity. If this were
removed, it would give an observed 240.72/246.20 ratio of 0.83$\pm$0.50, in agreement with theory (0.41),
albeit close to the limit of the error bar.

The observed 413.00/368.16 and 368.16/359.64 ratios are in good agreement with theory, indicating that the
relevant transitions are reliably detected and free from blends. In particular, there is no evidence that
the 368.16\,\AA\ feature is blended with a Cr\,{\sc xiii} line, as suggested by Thomas \& Neupert (1994). 
We note that the 
{\sc chianti} synthetic spectrum predicts that Cr\,{\sc xiii} should make less than a 1 per cent
contribution to the 368.16\,\AA\ intensity. 

The good agreement between theory and observation for both the 256.43/288.56 and 288.56/359.64
ratios provides support for our identification of the 288.56\,\AA\ line as being due to Fe\,{\sc xiii}, 
the first time (to our knowledge) this transition has been detected in the Sun.
Also, our results imply that that the 256.43\,\AA\ line is not blended with the
Zn\,{\sc xx} 3s $^{2}$S$_{1/2}$--3p $^{2}$P$_{3/2}$ transition, as suggested by both Thomas \& Neupert (1994)
and Keenan et al. (2003b). Indeed, Keenan et al. claim that, in the
SERTS--89 active region spectrum,
 Zn\,{\sc xx} is responsible for most of the measured line intensity.
However, this conclusion was based on the assumption that the 288.16\,\AA\ line in the SERTS--89 spectrum
is due to the Zn\,{\sc xx} 3s $^{2}$S$_{1/2}$--3p $^{2}$P$_{1/2}$ transition, 
as listed by Thomas \& Neupert.
If we adopt a solar Zn/Fe abundance ratio of 1.5$\times$10$^{-3}$ (Lodders 2003), and assume that other
atomic parameters are similar (a reasonable first approximation), we would expect the intensity of 
the Zn\,{\sc xx} 288.16\,\AA\ line to be 1.5$\times$10$^{-3}$ that of the isoelectronic 
Fe\,{\sc xvi} 360.75\,\AA\ feature, i.e. only  
8.0\,erg\,cm$^{-2}$\,s$^{-1}$sr$^{-1}$ (as opposed to the 
73.2\,erg\,cm$^{-2}$\,s$^{-1}$sr$^{-1}$ listed by Thomas \& Neupert).
In turn, this implies that the Zn\,{\sc xx} contribution to the 256.43\,\AA\ line is 
17.6\,erg\,cm$^{-2}$\,s$^{-1}$sr$^{-1}$, only 10 per cent of the total measured
line intensity. This is in agreement
with the present conclusions (and those of Landi 2002) that the 256.43\,\AA\ feature is not
significantly blended and is primarily due to Fe\,{\sc xiii}. An inspection of the {\sc chianti}
synthetic spectrum indicates that the 288.16\,\AA\ line is actually
due to the Ni\,{\sc xvi}
3s$^{2}$3p $^{2}$P$_{1/2}$--3s3p$^{2}$ $^{2}$D$_{3/2}$ transition. This is confirmed by the 
intensity ratio of this feature to Ni\,{\sc xvi} 239.49\,\AA, with the observed 
288.16/239.49 ratio = 0.34$\pm$0.19 (Thomas \& Neupert), 
compared to the theoretical value from {\sc chianti}
of 0.70. 

The observed 272.21/251.94, 272.21/256.43 and 272.21/359.64 ratios are all much larger
than theory,
indicating that the 272.21\,\AA\ feature is not due to Fe\,{\sc xiii}. Unfortunately,
our inspection of line lists reveals no likely candidate for 
this emission feature.

The observed 290.81/318.12 ratio is in reasonable agreement with theory, providing support for 
our new 
identification of the 290.81\,\AA\ feature as being due to Fe\,{\sc xiii}.
However the experimental ratio is somewhat larger than theory, indicating perhaps the presence of
some blending. On the other hand, we note that the {\sc chianti} synthetic spectrum, and other line lists,
do not indicate any suitable blending species. 

The observed and theoretical 311.57/320.80 ratios show no discrepancy, and indicate that both
lines are reliably measured and free from major blends. This is in contrast to the work of Landi (2002),
who suggested that the 311.57\,\AA\ feature is blended with a Cr\,{\sc xii}
line. Unfortunately, Cr\,{\sc xii} is not present in {\sc chianti}, but the synthetic spectrum does 
list a Fe\,{\sc ix} transition which will blend with 311.57\,\AA, although it is only predicted to 
contribute 12 per cent of the total line intensity.

The observed 321.46/312.17 and 321.46/348.18 ratios are in agreement
with the theoretical results within the error bars. However,
for 359.83/312.17 and 359.83/348.18 the experimental ratios
are somewhat smaller than theory, 
indicative of partial blending in both the
312.17 and 348.18\,\AA\ features, with the degree of blending more severe for the former.
This is in contrast to Landi (2002), who suggested that there are problems with the 321.46
and 359.83\,\AA\ lines. However, we find excellent agreement between theory and observation for
the 359.83/359.64 ratio, indicating no problem with 359.83\,\AA, although we note that there is 
a discrepancy with the {\sc chianti} calculation. An inspection of the {\sc chianti}
synthetic spectrum and other line lists reveals no likely 
blending candidate for the 312.17\,\AA\ transition, although there is an Fe\,{\sc ix} 
line which is predicted to contribute about 6 per cent to the total 348.18\,\AA\ intensity.
Hence we conclude that the blending of the 348.18\,\AA\ line, at least, is only minor.

There is excellent agreement between the observed and theoretical values of the 259.94/251.94 
ratio, confirming our new identification of the 259.94\,\AA\ transition in the solar spectrum.
However, the present theoretical line ratio is much larger than that from {\sc chianti}, although in this
instance the two sets of atomic data for the relevant transitions do not differ significantly.
For example, in the present calculations we use A = 
2.98$\times$10$^{8}$\,s$^{-1}$ and 3.56$\times$10$^{10}$\,s$^{-1}$ for the 259.94 and 251.94\,\AA\
transitions, respectively, from Table 5, while {\sc chianti} employs
A = 3.63$\times$10$^{8}$\,s$^{-1}$ and 3.37$\times$10$^{10}$\,s$^{-1}$ from Young (2004).
It is therefore unclear why there should be such large discrepancies in the theoretical ratios from
the two calculations. However we are confident that the present results are reliable.

The observed 312.89/320.80 ratio is in excellent agreement with theory, confirming the
identification of the 312.89\,\AA\ transition by Brickhouse, Raymond \& Smith (1995).
However the experimental value of 318.12/368.16 is much larger than the theoretical estimate, which
must be due to blending of the 318.12\,\AA\ feature as the 368.16/359.64 ratio shows good
agreement between theory and observation, confirming the reliability of the 368.16\,\AA\ detection.
Unfortunately, neither the {\sc chianti} synthetic spectrum nor other line lists indicate
suitable blending lines for the 318.12\,\AA\ feature.  
 
In summary, the problem Fe\,{\sc xiii} lines appear to be
240.72, 290.81, 312.17 and 318.12\,\AA\ in the SERTS--89 wavelength range, with the 
other 14 identified transitions being reliable measured, namely:
246.20, 251.94, 256.43, 259.94, 288.56, 311.57, 312.89, 
320.80, 321.46, 348.18, 359.64, 359.83, 368.16 and 413.00\,\AA. However the problem lines can all 
be explained either by blending, or by the fact that the line is weak and poorly observed. In particular, 
possible errors in the adopted atomic data do not need to be invoked to explain the discrepancies between
theory and observation for any of these lines.

\subsection{Electron density diagnostics}

In Tables 9 and 10 we summarise the observed values of electron density sensitive line 
intensity ratios for the SERTS--95 and SERTS--89 datasets, respectively, 
along with the derived log N$_{e}$
estimates. The densities have been determined from the present line ratio calculations at the
electron temperature of maximum Fe\,{\sc xiii} fractional abundance in ionization equilibrium, T$_{e}$
= 10$^{6.2}$\,K (Mazzotta et al. 1998), although we note that varying T$_{e}$ by $\pm$0.2 dex
(i.e. by 60 per cent) would change the derived
values of N$_{e}$ by typically $\pm$0.1 dex or less (see, for example, Fig. 6).
Also given in the tables is the factor by which the relevant ratio varies between N$_{e}$ = 10$^{8}$
and 10$^{11}$\,cm$^{-3}$, to show which are the most N$_{e}$--sensitive diagnostics.

\begin{table}
 \centering
\begin{minipage}{80mm}
  \caption{Electron density diagnostic line ratios from the SERTS 1995 active region 
spectrum.}
  \begin{tabular}{cccc}
  \hline
Line ratio &   Observed & log N$_{e}$$^{a}$ & Ratio variation$^{b}$ 
\\
\hline
196.53/202.04 & 0.11 $\pm$ 0.03 & 9.1$^{+0.1}_{-0.1}$ & 252
\\
200.03/202.04 & 0.25 $\pm$ 0.04 & 9.1$^{+0.1}_{-0.1}$ & 36
\\
201.13/202.04 & 0.39 $\pm$ 0.07 & 9.0$^{+0.1}_{-0.3}$ & 4.1
\\
201.13/203.79 & 1.5 $\pm$ 0.3 & 9.1$^{-0.2}_{+0.3}$ & 8.6
\\
203.17/202.04 & 0.13 $\pm$ 0.02 & 9.1$^{+0.1}_{-0.1}$ & 23
\\
203.83/202.04 & 1.3 $\pm$ 0.2 & 9.3$^{+0.1}_{-0.1}$ & 43
\\
209.63/202.04 & 0.17 $\pm$ 0.03 & 8.9$^{+0.1}_{-0.1}$ & 28
\\
209.91/208.67 & 5.3 $\pm$ 1.9 & 9.7$^{-0.2}_{+0.2}$ & 44
\\
\hline
\end{tabular}

$^{a}$Determined from present line ratio calculations at T$_{e}$ = 10$^{6.2}$\,K; N$_{e}$ in cm$^{-3}$. 
\\
$^{b}$Factor by which the theoretical line 
ratio varies between N$_{e}$ = 10$^{8}$ and 10$^{11}$\,cm$^{-3}$.
\end{minipage} 
\end{table}

\begin{table}
 \centering
\begin{minipage}{80mm}
  \caption{Electron density diagnostic line ratios from the SERTS 1989 active region 
spectrum.}
  \begin{tabular}{cccc}
  \hline
Line ratio &   Observed & log N$_{e}$$^{a}$ & Ratio variation$^{b}$ 
\\
\hline
259.94/318.12 & 0.45 $\pm$ 0.19 & 9.2$^{-0.4}_{+0.5}$ & 7.9
\\
290.81/348.18 & 0.039 $\pm$ 0.021 & 10.6$^{+\infty}_{-0.9}$ & 19
\\
311.57/348.18 & 0.24 $\pm$ 0.08 & 9.5$^{+0.4}_{-0.4}$ & 13
\\
312.17/320.80 & 0.49 $\pm$ 0.08 & 9.1$^{-0.1}_{+0.2}$ & 7.2
\\
312.17/368.16 & 0.63 $\pm$ 0.13 & 8.7$^{-0.2}_{+0.2}$ & 6.9
\\
312.89/312.17 & 0.55 $\pm$ 0.16 & 9.1$^{+0.6}_{-0.3}$ & 6.2
\\
312.89/348.18 & 0.36 $\pm$ 0.11 & 9.3$^{+0.4}_{-0.3}$ & 11
\\
318.12/320.80 & 0.51 $\pm$ 0.09 & 9.7$^{+0.2}_{-0.3}$ & 2.9$^{c}$
\\
318.12/348.18 & 0.69 $\pm$ 0.12 & 9.4$^{+0.1}_{-0.1}$ & 21
\\
320.80/251.94 & 0.48 $\pm$ 0.09 & 8.7$^{+0.3}_{-0.2}$ & 2.8
\\
321.46/368.16 & 0.25 $\pm$ 0.07 & 8.9$^{-0.2}_{+0.2}$ & 6.9
\\
348.18/256.43 & 0.90 $\pm$ 0.34 & 8.6$^{-0.6}_{+0.6}$ & 4.9
\\
348.18/320.80 & 0.74 $\pm$ 0.13 & 9.2$^{-0.1}_{+0.2}$ & 13
\\
348.18/368.16 & 0.95 $\pm$ 0.20 & 8.9$^{-0.1}_{+0.2}$ & 13
\\
359.64/348.18 & 1.2 $\pm$ 0.2 & 9.0$^{+0.2}_{-0.1}$ & 5.8
\\
359.83/318.12 & 0.26 $\pm$ 0.06 & 9.7$^{-0.2}_{+0.2}$ & 21
\\
359.83/368.16 & 0.17 $\pm$ 0.04 & 9.1$^{-0.1}_{+0.2}$ & 13
\\
413.00/348.18 & 0.056 $\pm$ 0.017 & 8.9$^{+0.2}_{-0.2}$ & 13
\\
\hline
\end{tabular}

$^{a}$Determined from present line ratio calculations at T$_{e}$ = 10$^{6.2}$\,K; N$_{e}$ in cm$^{-3}$. 
\\
$^{b}$Factor by which the theoretical line ratio varies between N$_{e}$ = 10$^{8}$ and 10$^{11}$\,cm$^{-3}$.
\\
$^{c}$Factor by which the theoretical line
ratio varies between N$_{e}$ = 10$^{9}$ and 10$^{11}$\,cm$^{-3}$ only 
(see
Fig. 7).
\end{minipage} 
\end{table}

As expected, for the SERTS--95 active region, the most consistent and reliable results in Table 9
come from the problem-free lines, with 196.53/202.04, 200.03/202.04, 201.13/203.79 and
203.17/202.04 
all yielding log $N_{e}$ = 9.1$\pm$0.1. Our recommendation is that 200.03/202.04 and 203.17/202.04
are preferentially employed as diagnostics, due to the wavelength
proximity of the transitions involved and the
N$_{e}$--sensitivity of the ratios. Although 196.53/202.04 is very density sensitive, the transitions
involved are further apart and the ratio is hence more
susceptible to possible errors in the instrumental
intensity calibration.  
We note that the average electron density found for the SERTS--95 active region in the present work 
using the subset of line ratios involving problem-free transitions, log N$_{e}$ = 9.1$\pm$0.1,
is very similar to that derived by Landi (2002) from the atomic data of Gupta \& Tayal (1998)
using the same set of diagnostics, namely log N$_{e}$ = 9.2$\pm$0.1.

Unfortunately, many of the density diagnostics 
in the SERTS--89 wavelength range are not particularly
N$_{e}$--sensitive, as may be seen from Table 10. As a result, 
most of the derived values of N$_{e}$ have 
very large error bars. However, the best diagnostics 
in terms of N$_{e}$--sensitivity, wavelength proximity of line pairs, and avoiding
problem lines, are judged to be 348.18/320.80, 348.18/368.16, 359.64/348.18 and 359.83/368.16. These
imply an average of 
log N$_{e}$ = 9.1$\pm$0.2, compared to log N$_{e}$ = 9.3$\pm$0.2 derived by Landi (2002)
from the same set of diagnostics. 

\section{Conclusions}

Our comparison of theoretical Fe\,{\sc xiii}
emission-line intensity
ratios with high resolution solar active region spectra from SERTS reveals generally
good agreement between theory and experiment, and has led to the identification of several new 
Fe\,{\sc xiii} emission features at 203.79, 259.94, 
288.56 and 290.81\,\AA. However, problems with several
Fe\,{\sc xiii}
transitions, first noted by Landi (2002) and Young et al. (1998), remain outstanding, and cannot
be explained by blending. Errors in the adopted atomic data appear to be the most likely
explanation, especially for transitions which have 
3s$^{2}$3p3d $^{1}$D$_{2}$ as their upper level, and
further calculations are urgently required.

For the SERTS--95 wavelength region (170--225\,\AA), we find that the 
ratios 200.03/202.04 and 203.17/202.04 provide the 
best Fe\,{\sc xiii}
density diagnostics, as they involve line pairs which appear to be problem-free, are close in wavelength and
are highly N$_{e}$--sensitive. Similarly, for the SERTS--89 range (235--450\,\AA) we recommend the use of
348.18/320.80, 348.18/368.16, 359.64/348.18 and 359.83/368.16 
to derive values of N$_{e}$ for the Fe\,{\sc xiii} emitting region of the plasma.

\section*{Acknowledgments}
KMA acknowledges financial support from EPSRC, while
DBJ is grateful to the Department of Education and Learning
(Northern Ireland) and NASA's Goddard Space Flight Center
for the award of a studentship.
The SERTS rocket programme is
supported by RTOP grants from the Solar Physics Office   
of NASA's Space Physics Division.                        
JWB acknowledges additional NASA support under
grant NAG5--13321.                                       
FPK is grateful to AWE Aldermaston for the award of a William Penney
Fellowship. The authors thank Peter van Hoof for the use of his
Atomic Line List, and Peter Young for extremely useful comments on an earlier version of this
paper.

\bsp

\label{lastpage}

\end{document}